\documentstyle[11pt,paspconf,epsf]{article}

\begin{document}
\title{A Stellar Library for Evolutionary Synthesis
       Modeling Including Variable AGB Stars.}

\author{M. Mouhcine, A. Lan\c{c}on}
\affil{Observatoire Astronomique de Strasbourg.\\ 
11, rue de l'Universit\'e, 67000, Strasbourg, France}

\begin{abstract}
A self-consistent spectrophotometric modelling of  
intermediate age post-starburst requires accurate stellar 
ingredients taking into account a principal feature of the 
stars dominating the near-IR emission during this phase: 
the variability of the AGB stars. A new library of stellar 
spectra based on averages of the empirical spectra 
of variable AGB stars is presented. This library is designed 
for convenient use in the population synthesis models. We 
discuss meaningful ways to compute these averages, and the 
non-trivial connection with the theoretical stellar parameters.  
Our library covers the near infrared wavelength range between 
$0.5\mu$m and $2.5\mu$m and exhibits fundamental differences 
when compared to the standard libraries using only static giants.

\end{abstract}


\section{Introduction}
Evolutionary population synthesis predictions 
depends strongly on the stellar inputs. 
One of them is the library of stellar spectra used to compute the 
integrated population spectrum. To interpret the integrated light 
of post-starburst galaxies in the near-IR, where the bulk of the energy 
of luminous evolved stars is emitted, 
one needs to introduce in some way the spectral signatures of such stars 
(essentially molecular absorption bands).
Asymptotic giant branch stars (AGB stars) are characterized by low 
temperatures as well as high luminosity.
The third dredge-up phenomenon on the AGB, due to the recurrent penetration 
of the convective envelope into the carbon-rich layers,
brings the products of the helium burning nucleosynthesis to the
surface (see Mowlavi, 1998, for a review), and
is responsible for the eventual conversion of some M stars to carbon stars. 
This process is highly efficient in metal-poor systems, where the 
C stars can dominate the AGB stars population luminosity at 
low metallicity (Groenewegen 1998). 
Another particularity of the most evolved AGB stars is 
their high rate of mass loss 
($\dot{M} \sim 10^{-7} - 10^{-4} M_{\odot}\,\, yr^{-1}$, Zijlstra 1998). 
These stars, surrounded by a dusty envelope, emit  most 
of their energy at wavelengths from a few to a few hundred microns. 
They can be missed in optical surveys, and accounting for circumstellar
extinction reduces the predicted emission of the AGB population
even in the near-IR spetral region (see Habing, 1996, for detailed 
review). 

For intermediate age populations one must
take into account all the AGB properties pointed out above. 
The different librairies used in the literature, empirical or 
synthetic ones, (Terndrup et al. 1990, Lan\c{c}on \& Rocca-Volmerange 
1992, Fluks et al. 1994, Lejeune et al. 1998) have introduced cool 
static giant stars, but they are not appropriate to distinguish 
between red giant stars and AGB stars. Bressan et al. (1998)
have taken the presence of the AGB stars into account 
in their modelling of the intermediate age population by
assigning, to each point of an isochrone, parameters of a
dusty envelope model and performing radiative transfer calculations 
to correct the spectra of static giant star. This method is promising 
but it takes into account only the effects of the circumstellar 
shells around AGB stars. Our principal motivation is to recognise 
the presence of AGB stars in the integrated near-IR spectrum of a stellar 
population and to introduce all the effects pointed out above to 
be able to interpret accurately the spectrophotometric propreties 
of intermediate age stellar populations. The scope of this paper 
is to construct a new stellar library taking into account the 
relevant processes affecting the near-IR spectra of AGB stars.    

\section{Data}
The spectra of our sample (Lan\c{c}on \& Wood, 1997)
cover a broad spectral range (0.5\,--\,2.5$\,\mu$m) 
quasi-instantaneously, i.e. with no phase mixing. 
This sample consists of a wide range of cool 
objects: $35$ O-rich Miras and $6$ C-rich Miras, with various 
periods (between $90$ and $450$ days, corresponding to different
luminosities), observed $3$ times or more on different phases 
(up to $8$ times). For reference and comparison, the sample also 
includes spectra of non-variable M-type giants, supergiants, 
LMC and galactic Bulge variables, and OH/IR and C/IR.
The near-IR spectra, taken at 2.3m ANU Telescope at Siding 
Spring Observatory, have a resolution of $1100$ and are 
connected with overlapping low resolution optical spectra, 
taken at Mt Stromlo Observatory. The quality of the spectra 
is good; the S/N per resolved element usually reaches 
$\sim 100$ in atmospheric windows and the typical uncertainty 
on (I-K) is $\leq 0.2$\,mag.\\
The data confirm that the long period variables display much
deeper near-IR molecular absorption bands than the cool
and luminous static stars (Bessell et al. 1996).
\begin{figure}
\plotfiddle{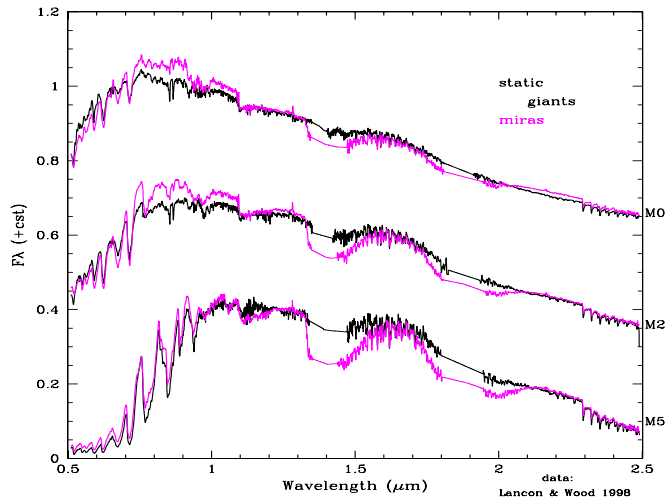}{4.5cm}{0}{45}{35}{-180}{-50}
\caption{Comparison between static giant and long period 
variable spectra. We see clearly how we can separate 
the two populations on the basis of the H$_{2}$O bands 
(1.4 and 1.9 $\mu$m) and how the difference is function 
of the spectral type (e.g. temperature). Dips around 
0.95 $\mu$m may be artifacts.} 
\end{figure}

\section{Constructing stellar library}
To estimate the luminosity of our spectra, we have used the 
L$_{K}$-P relation of Kanbur et al. (1998), and an instantanous 
bolometric correction, obtained from the individual spectra themselves
(Mouhcine et al. in preparation).\\
The effective temperature (T$_{\rm eff}$) 
determination is more problematic (see Haniff et al. 1995 for 
the essential reasons, which we cannot discuss in enough detail here),
and will strongly affect in the synthetic integrated spectra.
The shape of the spectral energy distribution of the O-rich 
long-period variables (LPVs) is similar at most phases to that of  
static M giants (Fig.\,1). Alvarez et al. (in preparation) 
find that the optical spectra of LPVs can be fitted well with
static giant model atmospheres. This might be interpreted as an
indication that the T$_{\rm eff}$ scale of the LPVs and static giants are 
equivalent. A large grid of static giant spectra was used to 
determine the effective temperatures of the spectra of our sample.
Feast (1996) has derived an alternative temperature scale for LPVs
using a data set grouping M Mira and non-Mira stars. The colour-T$_{\rm eff}$
relation derived is steeper than the one derived using the static giants 
models: cool Mira spectra are assigned lower T$_{\rm eff}$ values.
This relation could be biased since the T$_{\rm eff}$ used
in Feast's fit are derived from angular diameter measurements,
who are interpreted using uncertain models both to correct 
for limb-darkening and to convert a monochromatic radius 
to the effective radius (Rosseland optical depth 2/3 or 1.0). 
The situation is complicated by the uncertainties about the pulsation mode 
of some LPVs. In addition, the T$_{\rm eff}$ derived can be biased if 
there is some scattering source (dust?) in the upper atmospheres
of LPVs that makes them appear bigger than they are (Wood, private 
communication). \\
To be conservative we have constructed two different
librairies considering the two different T$_{\rm eff}$-(J-K) relations.

Schematically, the thermal pulses move stars up and down in luminosity
along the evolutionary tracks, while the LPV pulsations 
(periods of $10^2-10^3$ days) occur perpendicular to the tracks 
(large T$_{\rm eff}$ shifts), deeply modifying the   
spectral type, the shape of the global energy distribution 
and the spectral signatures of the stars. The evolutionary tracks 
represent the evolution of static parent stars of the pulsating 
long period variables. How do we deal with this problem?  How
to determine the stellar spectrum to associate with each point of the 
evolutionary tracks? There are two approaches to adopt. One is
to obtain properly weighted averages of the phase 
dependent spectra of individual stars, for various masses and 
evolutionary stages. This approach is, in principle, correct,
but needs a huge amount of data or large pulsating model grids, 
which makes this approach very hard practically.
The second approach, which was adopted, is to average the 
instantaneous spectra by temperature bin, disregarding phase
and the pulsation properties. This approach is supported by
the fact that, in our sample, no systematic correlation between 
the molecular indices and the luminosity or amplitude was found.
In addition, this approach needs much less data.\\
Our library also contains carbon stars. Using interferometric
angular diameters, in combinaison with bolometric flux, Dyck 
et al. (1997) have found that there is small range of effective 
temperatures for C-stars over a large range in spectral type, and 
derived that T$_{eff}=3000\pm200$ K. Finally,
the library includes 5 OH/IR and several observations of 1 C/IR star.    

\section{Conclusion}
This library of averaged spectra is constructed for population 
synthesis modelling. It should become public before the end of 
the year.
\acknowledgments 
We acknowledge the fruitful collaboration with R. Alvarez, 
B. Plez, M. Scholz, P. Wood, for different aspects of this work.

\end{document}